%% file: prl_dsds.tex
\documentclass[aps,prl,twocolumn,showpacs,groupedaddress,letterpaper,fleqn]{revtex4}  
\usepackage{graphicx}  
\usepackage{dcolumn}   
\usepackage{bm}        
\usepackage{amssymb}   
\usepackage{fleqn}

\newcommand{\ds}        {\ensuremath{D_s^{(\ast)}}}

\newcommand{\bs}        {\ensuremath{B_s^0}}
\newcommand{\brbs}        {\ensuremath{\mbox{Br}(\bs \to \ds \ds)}}
\newcommand{\pt}  {\ensuremath{p_{T}}}

\begin{document}

\hspace{5.2in} \mbox{FERMILAB-PUB-07/047-E}

\title{Measurement of the branching fraction Br$\bm{(\bs \to \ds \ds)}$}
\input list_of_authors_r2.tex  
\date{February 28, 2007}

\begin{abstract}
  We report a measurement of the branching fraction Br$(\bs \to \ds
  \ds)$ using a data sample corresponding to 1.3~fb$^{-1}$ of
  integrated luminosity collected by the D0 experiment in 2002--2006
  during Run~II of the Fermilab Tevatron Collider. One $\ds$ meson was
partially   reconstructed in the decay $D_s \to \phi \mu  \nu$, and the other
$\ds$ meson was identified using the decay $D_s \to \phi \pi$ 
  where no attempt was made to distinguish
$D_{s}$ and $D_{s}^{*}$ states. 
The resulting measurement is 
$\displaystyle{\brbs = 0.039^{+0.019}_{-0.017}\mbox{(stat)}
^{+0.016}_{-0.015}\mbox{(syst)}}$.
This was subsequently used to estimate the width difference $\Delta \Gamma^{CP}_{s}$ in the
  $\bs$--$\bar B^0_s$ system: $\Delta \Gamma^{CP}_{s} / \Gamma_{s}
  = 0.079^{+0.038}_{-0.035}\mbox{(stat)} ^{+0.031}_{-0.030}\mbox{(syst)}$.
\end{abstract}


\maketitle

In the standard model (SM), mixing in the \bs\ system is expected
to produce a large decay width difference $\Delta
\Gamma_{s}=\Gamma_{L}-\Gamma_{H}$ between the light and heavy mass eigenstates
with a small CP-violating phase $\phi_{s}$~\cite{dunietz}. New phenomena could
produce
a significant CP-violating phase leading to a reduction in the observed value
of $\Delta \Gamma_{s}$ compared with the SM prediction
of  $\Delta \Gamma_{s} / \Gamma_{s} = 0.127 \pm 0.024$~\cite{Lenz:2006hd}.
$\Delta \Gamma^{CP}_{s} = \Delta \Gamma_{s}^{\rm CP-even}-\Delta
\Gamma_{s}^{\rm CP-odd}$ $(\Delta \Gamma_{s} =\Delta\Gamma^{CP}_{s} \mid\cos
\phi_{s}\mid)$ can be
estimated from the branching fraction Br$(\bs \to \ds \ds)$~\cite{alexan,dunietz}.  This decay is
predominantly CP-even and is related to
$\Delta\Gamma^{CP}_{s}$~\cite{dunietz,alexan}:
$2\mbox{Br}(\bs \to \ds \ds) \approx \left(\Delta \Gamma^{CP}_{s}/{\Gamma_{s}}
\right) \left[ 1 + {\cal{O}}\left({\Delta \Gamma_{s}}/{\Gamma_{s}} \right)
\right] $.  
Only one measurement of \brbs~ has previously been published, by the 
ALEPH~\cite{aleph} experiment at the CERN LEP collider from the study of
correlated production of $\phi \phi$ in $Z^0$ decays.

In this Letter we present a measurement of \brbs\ using a
sample of semileptonic \bs\ decays collected by the D0 
experiment at Fermilab in $p\bar{p}$ collisions at $\sqrt{s} =
1.96$~TeV. The data correspond to an integrated luminosity of
approximately 1.3~fb$^{-1}$.  
We present an analysis of the decay chain
$\bs \to \ds \ds$ where one $D_s^{+}$ decays to $\phi_{1} \pi^{+}$, the other
 $D_s^{-}$ decays to $D_s^{-} \to  \phi_{2}\mu^{-} \nu$, and where each
$\phi$ meson decays to $\phi \to K^+ K^-$. Charge conjugate states are implied
throughout.
No attempt was made to reconstruct the photon or $\pi^{0}$ from the decay
$D_{s}^{*}\to D_{s}\, \gamma/ \pi^{0}$ and thus the state $\ds\ds$ contains
contributions from $D_{s}D_{s}$, $D_{s}^{*} D_{s}$ and $D^{*}_{s} D^{*}_{s}$.
To reduce systematic effects, $\brbs$ was normalized to
the decay $\bs\to\ds\mu\nu$.

The D0 detector is described in detail elsewhere~\cite{run2det}. The
detector components relevant to this analysis are the central
tracking and muon systems. The D0 central-tracking system consists
of a silicon microstrip tracker (SMT) closest to the beampipe surrounded by a
central scintillating-fiber tracker (CFT) with an outer radius of $52$ cm. Both
tracking systems are
located within a $2$~T superconducting solenoidal magnet
and are optimized for tracking and vertexing for pseudorapidities 
$|\eta|<3$ (SMT) and $|\eta|<2.5$ (CFT), where $\eta$ =
$-$ln[tan($\theta$/2)], and $\theta$ is the polar angle with respect to the
beam axis.
The muon system
is located outside of the liquid-argon/uranium calorimetry system and has
pseudorapidity coverage $|\eta|<2$. 
It consists of a layer of tracking detectors and trigger scintillation counters
in front of a $1.8$~T iron toroid, followed by two similar layers outside of the
toroid. 
The trigger system identifies events of interest in a high-luminosity environment 
based on muon identification, charged tracking, and
vertexing. 
No explicit trigger requirement was applied, however most events satisfied
inclusive single-muon triggers.

 The measurement began with reconstruction of the decay chain 
$D_s \to \phi_{1} \pi$, $\phi_{1} \to K^+ K^-$, with tracks originating from the
same $p\bar{p}$ collision point (primary vertex) as a muon.
 All charged tracks used in
the analysis were required to have at least two hits in both the SMT and CFT.
Muons were required to have
transverse momentum $\pt >2$~GeV$/c$, total momentum $p>3$~GeV$/c$,
and to have measurements in at least two layers of the muon system.
Two oppositely charged particles with $\pt > 0.8$ GeV/$c$ were
selected from the remaining particles in the event and were assigned
the mass of a kaon.  An invariant mass of $1.01 < M(K^+ K^-)<1.03$ GeV/$c^{2}$
was required, 
to be consistent with the mass of a $\phi$ meson. 
Each pair of kaons satisfying these criteria was combined with a third particle
with $\pt > 1.0$
GeV/$c$, which was assigned the mass of a pion. 
 The three tracks were required
to form a
$D_s$ vertex using the algorithm described in Ref.~\cite{PVref}. 
The cosine of the angle between
the $D_s$ momentum and the direction from the primary vertex
to the $D_s$ vertex was required to be greater than $0.9$.
The $D_s$ vertex was required to have a displacement from the  
primary vertex in the plane
perpendicular to the beam with at least 4$\sigma$ significance.  
The helicity angle $\chi$ is defined as the angle between the momenta of the
$D_{s}$ and a $K$ meson in the $(K^+ K^-)$ center of mass system. 
The decay of $D_{s}\to \phi \pi$ follows a $\cos^{2}\chi$ distribution,
while for background $\cos\chi$ is expected to be flat.  Therefore, to enhance
the signal, the criterion $|\cos\chi|>0.35$ was applied.
 The muon and pion were required to have opposite charge.
The events passing these selections, referred to as the preselection sample, were
used to produce the samples of  $(\mu \phi_{2} D_{s})$ and the normalizing
sample $(\mu D_{s})$ defined below.

\begin{figure}
  \begin{center}
    \includegraphics[width=0.49\columnwidth]{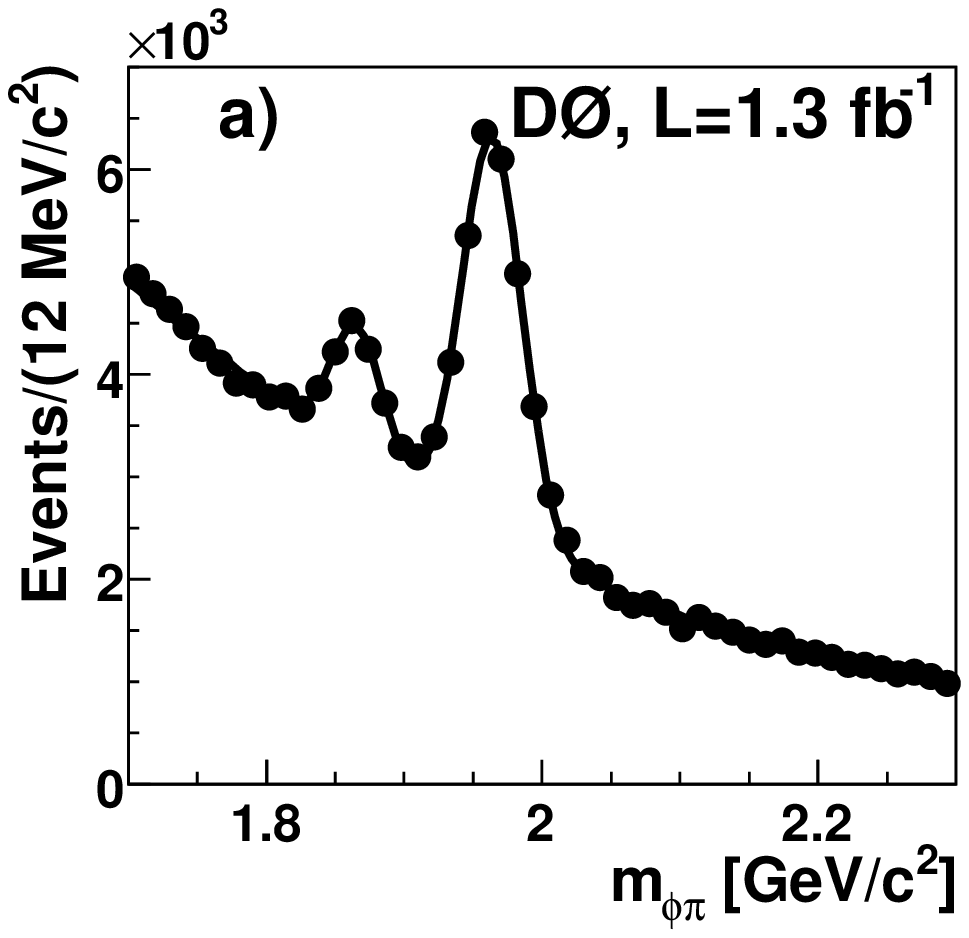}
    \includegraphics[width=0.49\columnwidth]{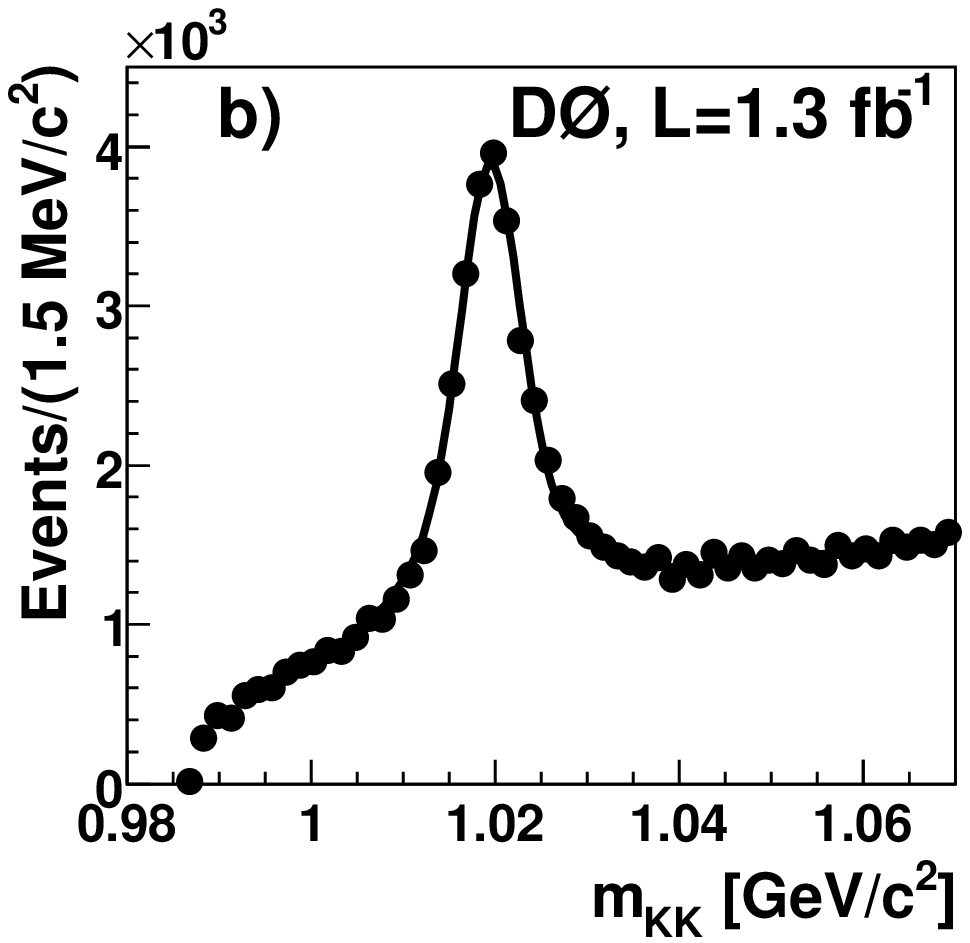}
    \caption{\label{fig:dsmu_data_kkpi}
     (a)~The  ($K^+ K^- \pi$) invariant
mass spectrum in the mass window $1.01 < M(K^+ K^-) < 1.03$~GeV/$c^2$. 
    (b)~Mass spectrum of the ($K^+ K^-$) system in the mass window 
$1.92 < M(K^+ K^- \pi) < 2.00$~GeV/$c^2$.
 }
  \end{center}
\end{figure}

To construct a $(\mu D_{s})$ candidate from the preselection sample,
the $D_{s}$ candidate and the
muon were required to originate from a common  \bs\ vertex. 
The mass of the $(\mu D_{s})$ system was required to
be less than 5.2 GeV/$c^2$. 
The number of tracks near the $\bs$ meson tends to be small, thus to 
reduce the background from combinatorics, an isolation criterion was applied.
The isolation is defined as the sum of the momenta of the tracks used to
reconstruct the signal divided
by the total momentum of tracks contained within a cone of radius $\Delta
{\cal R} = \sqrt{\Delta \eta ^2 + \Delta \phi^2} = 0.5$ centered on the direction
of the $\bs$ candidate. We required the isolation to exceed $0.6$. 
To suppress background, the visible proper decay length, defined as 
$M(\bs)\cdot( {\vec L_{T}} \cdot {\vec p_{T}} )/p^{2}_{T}$,
was required to exceed $150$~$\mu$m.
Here ${\vec L_{T}}$ is the displacement from the primary vertex to the $\bs$
decay vertex in the transverse plane,
and $M(\bs)$ is the mass of the $\bs$ meson~\cite{pdg}.
These data are referred to as the ($\mu D_s$) sample; the resulting mass
spectrum of the ($K^+ K^- \pi$) system is shown in
Fig.~\ref{fig:dsmu_data_kkpi}(a), where the $D_{s}$ and $D^{+}$ mass peaks 
are described by single Gaussians with a second-order polynomial to
parameterize the background.
The signals of $D_s \to \phi_{1} \pi$ and $D^{+} \to \phi_{1} \pi^{+}$ are
clearly seen.
Figure~\ref{fig:dsmu_data_kkpi}(b) shows the mass spectrum of the ($K^+ K^-$)
system, 
%
where a double Gaussian describes the $\phi$ mass peak, and a second-order
polynomial is used to parameterize the background.

To construct a ($\mu \phi_{2} D_s$) candidate from the preselection sample,
a second $\phi_{2}$ meson from $D_{s}\to\phi_{2}\mu\nu$ was required.
The selection criteria to reconstruct the second $\phi_{2}$ meson
were identical to those of the first $\phi_{1}$ meson, with the exception
that a wider mass range $0.99 < M(K^+ K^-) < 1.07$ GeV/$c^2$ was used
to estimate the background distribution under the $\phi_{2}$ meson.  
This $\phi_{2}$ meson and muon were required to form a  $D_{s}$ vertex.
To suppress background, the mass of the
$(\mu \phi_{2})$ system was required to be $1.2 < M(\mu \phi_{2}) < 1.85$
GeV/$c^2$. The $D_s(\phi_{1} \pi)$ and $D_s(\phi_{2} \mu)$ mesons were required
to form a $\bs$ vertex. The mass of the $(\mu \phi_{2} D_s)$ system, i.e., the
combined $D_{s}\to \phi_{2}\mu\nu$ and $D_{s}\to\phi_{1}\pi$ candidates,
 was required to be $4.3 < M(\mu \phi_{2} D_s) < 5.2$ GeV/$c^{2}$.  
An isolation value exceeding $0.6$ and visible proper decay length
greater than $150$~$\mu$m were required for the $\bs$ meson.

To reduce the effect of systematic uncertainties, we calculated the
ratio $R = \brbs \cdot\mbox{Br}(D_s \to \phi \mu \nu)/\mbox{Br}(\bs\to\mu \nu
\ds)$. We extracted \brbs\
from $R$ using the known values~\cite{pdg} for Br$(D_s \to \phi \mu
\nu)$, Br$(\bs \to \ds\mu \nu )$, and Br$(D_s \to \phi \pi)$. $R$ can
be expressed in terms of experimental observables:\\
\begin{eqnarray}
\label{eq:major}
  \displaystyle{\!\!\!\!\!\!\!\!\!\!R} &= &
  \displaystyle{\frac{N_{\mu \phi_{2} D_s} - N_{\rm bkg}}
     {N_{\mu D_s}~f(\bs \to \ds\mu \nu )} \nonumber \times }\\
   &&\displaystyle{  \frac{1}{2\mbox{Br}(\phi \to K^+ K^-)}
  \frac  {\varepsilon(\bs \to \ds\mu \nu )}
             {\varepsilon(\bs \to \ds \ds)},}
\end{eqnarray}
where $N_{\mu D_s}$ is the number of ($\mu D_s$) events, $N_{\mu \phi_{2}
  D_s}$ is the number of ($\mu \phi_{2} D_s$) events, $N_{\rm bkg}$ is the
number of background events in the ($\mu \phi_{2} D_{s}$) sample that are
not produced by $\bs \to \ds \ds$ decays, and $f(\bs \to \ds\mu \nu)$ is the
fraction of events in ($\mu D_s$) coming from $\bs \to \ds \mu \nu X$.
The ratio of efficiencies  $\varepsilon(\bs \to \ds \ds)/
\varepsilon(\bs \to\ds\mu\nu)$ to reconstruct the two processes was determined
from simulation.
All processes involving $b$ hadrons were simulated 
with {\sc EvtGen}~\cite{EvtGen}
interfaced to {\sc pythia}~\cite{pythia},
followed by full modeling of the detector response with 
{\sc geant}~\cite{geant} and event reconstruction as in data.
The number of ($\mu D_s$) events was estimated from a binned fit to the
($K^+ K^- \pi$) mass distribution shown in
Fig.~\ref{fig:dsmu_data_kkpi}(a) from the $145,\!000$ candidates passing the
selection criteria.  
The resulting fit is superimposed in
Fig.~\ref{fig:dsmu_data_kkpi}(a) as a solid line and gives $N_{\mu
  D_s} = 17670 \pm 230~\mbox{(stat)}$ events.

The number  of ($\mu \phi_{2} D_s$) events was extracted using a
unbinned log-likelihood fit to the two-dimensional distribution of the
invariant masses $M_D$ of the ($\phi_{1} \pi$) system and
 $M_{\phi_{2}}$ of the two additional kaons from the $(\phi_{2}\mu)$ system.
All candidates from the ($\mu \phi_{2} D_s$) sample with $1.7 < M_D < 2.3$ GeV/$c^2$
and $0.99 < M_{\phi_{2}} < 1.07$ GeV/$c^2$ were included in the fit.
In the fit, the masses and widths for both $D_s$ and $\phi$ signals were
fixed to the values extracted from a fit to the ($\mu D_{s}$) data
sample. 
Extracted from the fit were the numbers of:
$N_{\mu \phi_{2} D_s}$ events from correlated (joint) signal
production of $(\phi_{1}\pi)$ and $\phi_{2}$,
events with a reconstructed $(\phi_{1}\pi)$ in the mass peak of
$D_{s}(\phi_{1}\pi)$ without joint production of $\phi_{2}$ from
$(\phi_{2}\mu)$ (i.e., uncorrelated),
events with a reconstructed $\phi_{2}$ from $(\phi_{2}\mu)$ without joint
production of $(\phi_{1}\pi)$ in the mass peak of the $D_{s}(\phi_{1}\pi)$
(i.e., also uncorrelated),
and combinatorial background.

\begin{figure}
  \begin{center}
    \includegraphics[width=0.49\columnwidth, height=0.65\columnwidth]{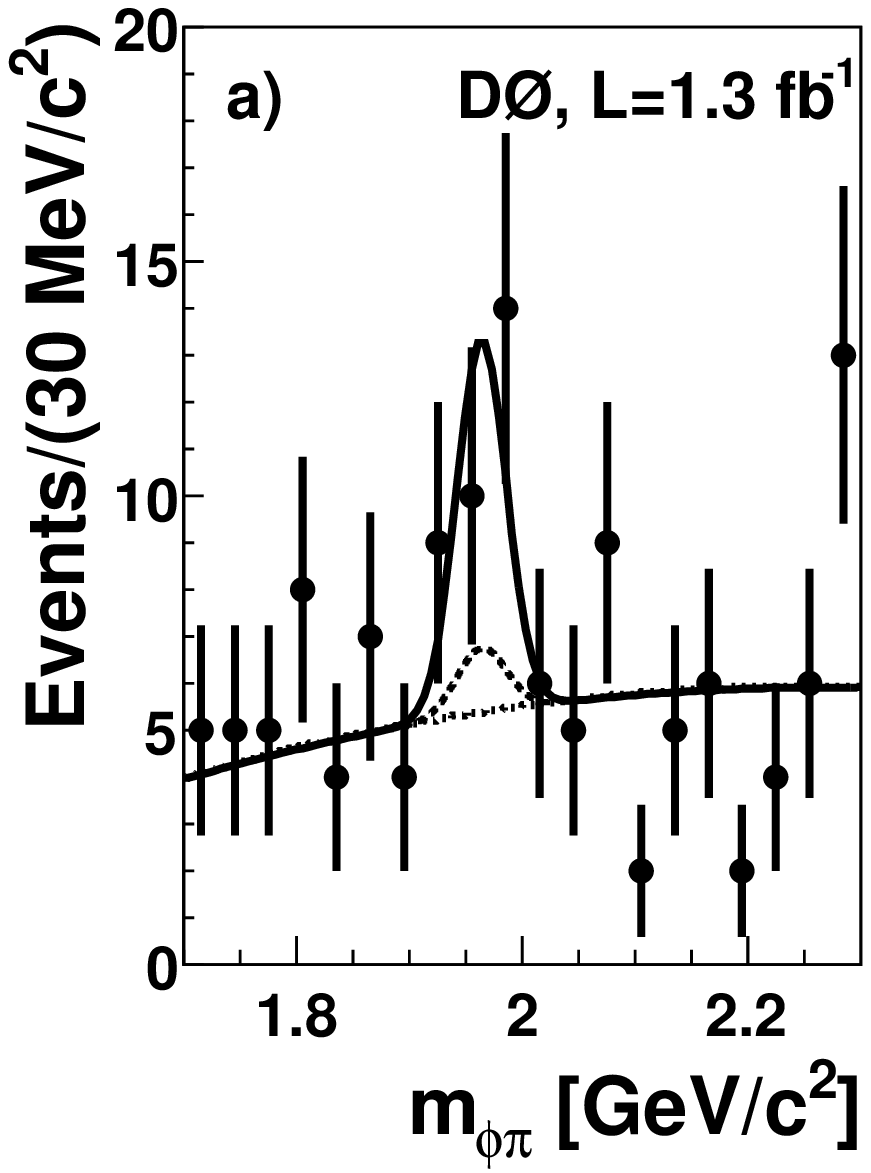}
    \includegraphics[width=0.49\columnwidth, height=0.65\columnwidth]{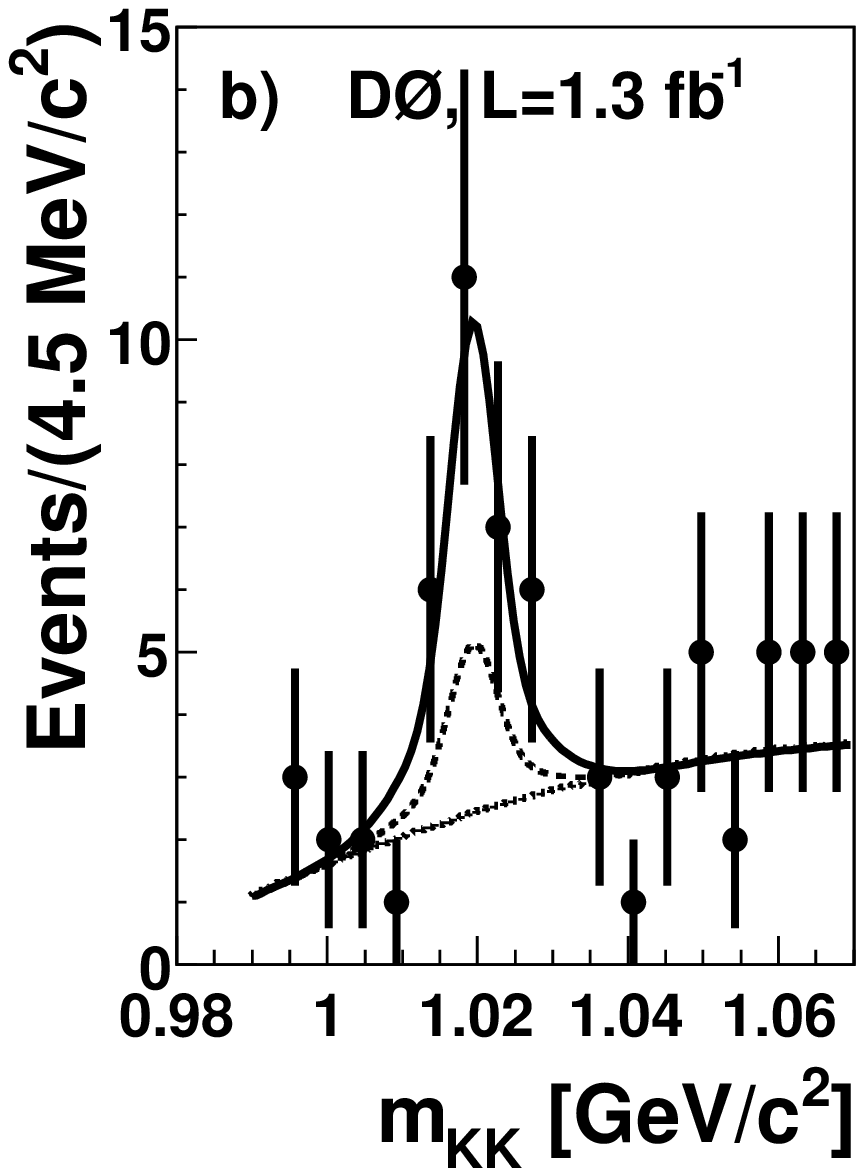}
    \caption{\label{fig:dsds_data_llfit_ds} Invariant mass distributions of
(a)~$D_{s}(\phi_{1}\pi)$ events in the signal
window of the invariant mass $(K^{+}K^{-})$ from $D_{s}(\phi_{2}\mu)$, and 
(b)~$(K^{+}K^{-})$ events from 
$D_{s}(\phi_{2}\mu)$ in  the invariant mass window of the  $D_{s}(\phi_{1}\pi)$
signal region. The solid curve is the projected result of the unbinned log-likelihood
fit, the dotted curve shows the polynomial background contribution, and the
dashed line shows the uncorrelated production of  (a)~$D_{s}(\phi_{1}\pi)$ and
(b)~$\phi_{2}$ mesons.}
  \end{center}
\end{figure}

The results of the fit are displayed in Fig.~\ref{fig:dsds_data_llfit_ds}.  
Figure~\ref{fig:dsds_data_llfit_ds}(a) displays the invariant mass
distribution of ($\phi_{1}\pi$)
candidates from the invariant mass signal window of $D_{s}(\phi_{2}\mu)$, and
Fig.~\ref{fig:dsds_data_llfit_ds}(b) displays the $\phi_{2}$ meson from
$D_{s}(\phi_{2}\mu)$ in the invariant mass signal decay window of
$D_{s}(\phi_{1}\pi)$ candidates.
The fit gives $N_{\mu \phi_{2} D_s} = 13.4^{+6.6}_{-6.0}$ events from the 340
candidates included in the fit.

As a consistency check, a similar sample was produced, but requiring the
same charge for the muon and pion. From the fit, the number of 
$(\mu^{+}\phi_{2} D_{s}^{+})$ signal events is found to be zero with 
an upper limit of $2.6$ events ($68\%$ CL).

To extract the number of $\bs \to \ds\mu \nu $ and $\bs \to \ds \ds$
events, the composition of the selected samples must be
determined. 
The decays $\bs \to \ds \mu \nu  X$ and 
$\bs \to \ds \tau(\to\mu\nu) \nu X$
were considered as signal.
The branching fractions for $B \to D_s D^{(*)} X$ and $\bs \to \ds \ds$ are
taken from the Ref.~\cite{pdg}. There is no experimental information
for the Br($\bs \to D_s D X$), therefore we used the value 15.4\% provided 
by Ref.~\cite{EvtGen} with an assigned uncertainty of $100\%$.

In addition, the ($\mu D_s$) sample includes the processes $c \bar{c}
\to D_s \mu \nu X$, $b \bar b \to  D_s \mu \nu X$, and events with a
misidentified muon, etc. 
which we refer to as the ``peaking background.''  
The estimated contribution of 
these processes in the ($\mu D_s$)
signal is $(2 \pm 1)$\%.  In total, we estimate that the
fraction of events in the ($\mu D_s$) signal coming from $\bs \to \ds\mu \nu  X$
is $f(\bs \to \ds\mu \nu ) = 0.82 \pm 0.05$.

We considered the number of events $N_{\mu \phi_{2} D_s}$ from the ($\mu
\phi_{2}D_s$)
sample to contain contributions from  1)~the main signal $\bs\to\ds\ds$,  and
the following background processes
2)~$B \to \ds \ds K X$,
3)~$\bs \to \ds \ds X$,
4)~$\bs \to  \ds \phi \mu \nu $,
5)~peaking background, and
6)~$\bs \to \ds \mu \nu $ combined with a $\phi$ meson from fragmentation.
There is no experimental information for most
of the processes, therefore their contributions were estimated by
counting events in different regions of the ($\mu \phi_{2} D_s$) phase
space and comparing the obtained numbers with the expected mass
distribution for each background process.

The mass of the ($\mu \phi_{2} D_s$) system for the second and third
processes is much less than that for the main decay $\bs \to \ds \ds$
because of the additional particles,
and the requirement $M(\mu \phi_{2} D_s) > 4.3$ GeV/$c^2$ strongly suppresses
them.  The contribution of $\bs \to \ds \ds X$ is much less
than $B \to \ds \ds K X$ because of higher production rates of $B^+$
and $B^0$ compared to \bs. 
Compared to the $B \to \ds \ds K X$ process, the final state in the decay 
 $\bs \to \ds \ds X$ includes at least two pions due to isospin
considerations. At least two gluons are required to produce this state
(similar to $\psi(2S) \to J/\psi \pi \pi$); it is therefore
additionally suppressed and its contribution was neglected. 
Simulation shows that for the $B \to \ds \ds K X$ decay, the
fraction of events with $ M(\mu \phi_{2} D_s) > 4.3$~GeV/$c^{2}$ is 0.05.
Requiring $M(\mu \phi_{2} D_s) < 4.3$ GeV/$c^2$ and keeping all
other selections, we observe $2.8 ^{+11.2}_{-2.8}$ events  in data.  
Assuming that all these events are due to $B \to \ds
\ds K X$, we estimate their contribution to the signal ($\mu \phi_{2}
D_s$) as $0.14^{+0.56}_{-0.14}$ events.

The fourth process produces a high mass for both the ($\mu
\phi_{2}$) and ($\mu \phi_{2} D_s$) systems and requiring $M(\mu \phi_{2}) <
1.85$ GeV/$c^2$ strongly suppresses it. Simulation shows that
for this process, the fraction of events with $M(\mu \phi_{2}) <
1.85~\mbox{GeV}/c^2$ is 0.14.  Requiring
 $M(\mu \phi_{2}) > 1.85$ GeV/$c^2$ and keeping all other selections, we observe
$13 \pm 11$ events. 
Assuming that all these events are due to
the fourth background process, we estimate its contribution to the
($\mu \phi_{2} D_s$) signal as $1.88 \pm 1.51$ events.

The contribution of the peaking background is strongly suppressed
by the event selection  with an upper limit of $0.4$ events. We therefore 
included it as an additional uncertainty in the number of background events.

The fitting procedure accounts for the possible background contribution of the
decay $\bs \to \ds\mu \nu $ together with the uncorrelated production 
of a $\phi$ meson from fragmentation.
In addition, an attempt was made to reconstruct ($\mu \phi_{2}
D_s$) events in the $\bs \to \ds\mu \nu $ simulation 
containing approximately $9200$ reconstructed ($\mu D_s$) events, and no such
events were found. Therefore the contribution of this process was
neglected.
In total, we estimate the number of background events as $N_{\rm bkg}
= 2.0 \pm 1.6$.

In determination of efficiencies,
the final states in the $(\mu D_{s})$ and $(\mu \phi_{2} D_{s})$ samples differ
only by the two kaons from the additional $\phi_{2}$ meson. With the exception
of the isolation criterion, all other applied selections are the same, so many
detector-related systematic uncertainties cancel. 
The muon \pt\ spectrum in $\bs \to \ds \mu \nu $
decay differs between data and simulation due to trigger effects and the
uncertainties in $B$ meson production in simulation.  
To correct for this difference, weighting functions were applied to all Monte
Carlo events. They were obtained from the ratio of simulated and data events 
for $\pt$ distributions of the $\bs$ meson and muon.
With this correction, the ratio of efficiencies is
 ${\varepsilon(\bs \to \ds \ds)}/ {\varepsilon(\bs \to \ds\mu \nu )} =
0.055\pm$0.001~\mbox{(stat)}.
The systematic uncertainty in this value is discussed below.
The difference in efficiency is mainly due to the 
softer momentum spectrum of the muon from the $D_{s}\to\phi\mu\nu$ decay in the
$(\mu\phi_{2}\ds)$ sample, compared to the muon from the $\bs\to\ds\mu\nu$ decay
in the $(\mu\ds)$ sample.

Using all these inputs and taking the value Br$(\phi \to K^+ K^-) = 0.492 \pm
0.006$~\cite{pdg}, we obtain $R = 0.015 \pm
0.007~\mbox{(stat)}$. The statistical uncertainty
shown includes only the uncertainty in $N_{\mu \phi_{2} D_s}$. All other
uncertainties are included in the systematics.

The experimental extraction of both  $\mbox{Br}(\bs \to \ds \mu \nu)$ 
and $\mbox{Br}(D_s \to \phi \mu\nu)$ depend on $\mbox{Br}(D_s \to \phi \pi)$.
Factorizing the
dependence on $\mbox{Br}(D_s \to \phi \pi)$, we obtain $\mbox{Br}(\bs
\to \mu \nu \ds) \mbox{Br}(D_s \to \phi \pi) =
 (2.84 \pm 0.49)\times 10^{-3}, \mbox{Br}(D_s \to \phi \mu \nu) =
(0.55 \pm 0.04)\cdot\mbox{Br}(D_s \to \phi \pi)$.  Using these numbers,
we finally obtain $\brbs = 0.039 ^{+0.019}_{-0.018}\mbox{(stat)}$.

The systematic uncertainties in the measured value of \brbs~ were
estimated as follows. All external branching fractions~\cite{pdg}
were varied within one standard deviation. A 100\% uncertainty in the
number of background events $N_{\rm bkg}$ in the ($\mu \phi_{2} D_s)$
sample was assumed. The ratio of efficiencies can be affected by the
uncertainties of reconstruction of two additional charged particles
from the $\phi$ meson decay. 
The efficiency to reconstruct a charged pion from
the decay $D^{*+} \to D^0 \pi^+$ was measured in Ref.~\cite{brat},
and the obtained value was in a good agreement
with the MC estimate.  This comparison is valid within the uncertainty
of branching fractions of different $B$ meson semileptonic decays, which is
about 7\%. Therefore we conservatively assigned a 14\% systematic
uncertainty (7\% for each charged particle, 100\% correlated) to the
ratio of efficiencies and propagated it to the final result.  For the
ratio of efficiencies, a 15\% uncertainty was assigned for the reweighting
procedure, which reflects the difference in efficiency between
weighted and unweighted estimates. The dependence of the number of
($\mu \phi_{2} D_s$) events on the fitting procedure was estimated by adding a
possible signal contribution from $D^+$ events which decreased the correlated
signal by 3\%, which we assigned as a systematic uncertainty.

Using these numbers, we obtain $\brbs = \left[0.039^{+0.019}_{-0.017}\mbox{(stat)}
\pm 0.014\mbox{(syst)}\right] \cdot \left[{0.044}/{\mbox{Br}(D_s \to \phi \pi)}
\right]^2$.  Using $\mbox{Br}(D_s \to \phi \pi) = 0.044 \pm 0.006$~\cite{pdg},
we find
\begin{equation}
\!\!\!\!\!\!\!\!\!\!\!\!\!
\brbs = 0.039^{+0.019}_{-0.017}\mbox{(stat)}^{+0.016}_{-0.015}\mbox{(syst)}. 
\end{equation}
The result is consistent with, and more precise than the ALEPH measurement
$ \brbs = 0.077 \pm 0.034 ^{+0.038} _{-0.026}$~\cite{aleph}, where the value
has been recalculated using the current value of 
$\mbox{Br}(D_s \to \phi \pi)$~\cite{pdg}.
We calculate $\Delta \Gamma^{CP}_{s}$~\cite{dunietz}
assuming that the decay $\bs \to \ds \ds$ is mainly CP-even and gives the 
primary contribution to the width difference between the CP-even and CP-odd 
$\bs$ states~\cite{alexan}:
\begin{equation}
\frac{\Delta \Gamma^{CP}_{s}}{\Gamma_{s}} = 0.079^{+0.038}_{-0.035}\mbox{(stat)}
^{+0.031}_{-0.030}\mbox{(syst)}.
\end{equation}
Assuming CP-violation in $\bs$ mixing is small~\cite{Lenz:2006hd}, this
estimate is in good agreement with the SM
prediction $\Delta \Gamma_{s} / \Gamma_{s} = 0.127 \pm 0.024$~\cite{Lenz:2006hd} 
and with the direct measurement of this parameter by
the D0 experiment in $\bs \to J/\psi \phi$ decays \cite{d0-dg}.  The
agreement with the CDF measurement of $\Delta \Gamma_{s} / \Gamma_{s}$,
which was also performed in $\bs \to J/\psi \phi$~\cite{cdf-dg}, is
not as good, although still within two standard deviations.

\input acknowledgement_paragraph_r2.tex

\end{document}

%% file: list_of_authors_r2.tex
%
\author{                                                                      
V.M.~Abazov,$^{35}$                                                           
B.~Abbott,$^{75}$                                                             
M.~Abolins,$^{65}$                                                            
B.S.~Acharya,$^{28}$                                                          
M.~Adams,$^{51}$                                                              
T.~Adams,$^{49}$                                                              
E.~Aguilo,$^{5}$                                                              
S.H.~Ahn,$^{30}$                                                              
M.~Ahsan,$^{59}$                                                              
G.D.~Alexeev,$^{35}$
% ...
\\                                                                            
\vskip 0.30cm                                                                 
\centerline{(D\O\ Collaboration)}                                             
\vskip 0.30cm                                                                 
}                                                                             
\affiliation{                                                                 
\centerline{$^{1}$Universidad de Buenos Aires, Buenos Aires, Argentina}       
\centerline{$^{2}$LAFEX, Centro Brasileiro de Pesquisas F{\'\i}sicas,         
                  Rio de Janeiro, Brazil}                                     
\centerline{$^{3}$Universidade do Estado do Rio de Janeiro,                   
                  Rio de Janeiro, Brazil}                                     
\centerline{$^{4}$Instituto de F\'{\i}sica Te\'orica, Universidade            
                  Estadual Paulista, S\~ao Paulo, Brazil}                     
\centerline{$^{5}$University of Alberta, Edmonton, Alberta, Canada,           
                  Simon Fraser University, Burnaby, British Columbia, Canada,}
\centerline{York University, Toronto, Ontario, Canada, and                    
                  McGill University, Montreal, Quebec, Canada}                
\centerline{$^{6}$University of Science and Technology of China, Hefei,       
                  People's Republic of China}                                 
\centerline{$^{7}$Universidad de los Andes, Bogot\'{a}, Colombia}             
\centerline{$^{8}$Center for Particle Physics, Charles University,            
                  Prague, Czech Republic}                                     
\centerline{$^{9}$Czech Technical University, Prague, Czech Republic}         
\centerline{$^{10}$Center for Particle Physics, Institute of Physics,         
                   Academy of Sciences of the Czech Republic,                 
                   Prague, Czech Republic}                                    
\centerline{$^{11}$Universidad San Francisco de Quito, Quito, Ecuador}        
\centerline{$^{12}$Laboratoire de Physique Corpusculaire, IN2P3-CNRS,         
                   Universit\'e Blaise Pascal, Clermont-Ferrand, France}      
\centerline{$^{13}$Laboratoire de Physique Subatomique et de Cosmologie,      
                   IN2P3-CNRS, Universite de Grenoble 1, Grenoble, France}    
\centerline{$^{14}$CPPM, IN2P3-CNRS, Universit\'e de la M\'editerran\'ee,     
                   Marseille, France}                                         
\centerline{$^{15}$Laboratoire de l'Acc\'el\'erateur Lin\'eaire,              
                   IN2P3-CNRS et Universit\'e Paris-Sud, Orsay, France}       
\centerline{$^{16}$LPNHE, IN2P3-CNRS, Universit\'es Paris VI and VII,         
                   Paris, France}                                             
\centerline{$^{17}$DAPNIA/Service de Physique des Particules, CEA, Saclay,    
                   France}                                                    
\centerline{$^{18}$IPHC, IN2P3-CNRS, Universit\'e Louis Pasteur, Strasbourg,  
                   France, and Universit\'e de Haute Alsace,                  
                   Mulhouse, France}                                          
\centerline{$^{19}$IPNL, Universit\'e Lyon 1, CNRS/IN2P3, Villeurbanne, France
                   and Universit\'e de Lyon, Lyon, France}                    
\centerline{$^{20}$III. Physikalisches Institut A, RWTH Aachen,               
                   Aachen, Germany}                                           
\centerline{$^{21}$Physikalisches Institut, Universit{\"a}t Bonn,             
                   Bonn, Germany}                                             
\centerline{$^{22}$Physikalisches Institut, Universit{\"a}t Freiburg,         
                   Freiburg, Germany}                                         
\centerline{$^{23}$Institut f{\"u}r Physik, Universit{\"a}t Mainz,            
                   Mainz, Germany}                                            
\centerline{$^{24}$Ludwig-Maximilians-Universit{\"a}t M{\"u}nchen,            
                   M{\"u}nchen, Germany}                                      
\centerline{$^{25}$Fachbereich Physik, University of Wuppertal,               
                   Wuppertal, Germany}                                        
\centerline{$^{26}$Panjab University, Chandigarh, India}                      
\centerline{$^{27}$Delhi University, Delhi, India}                            
\centerline{$^{28}$Tata Institute of Fundamental Research, Mumbai, India}     
\centerline{$^{29}$University College Dublin, Dublin, Ireland}                
\centerline{$^{30}$Korea Detector Laboratory, Korea University,               
                   Seoul, Korea}                                              
\centerline{$^{31}$SungKyunKwan University, Suwon, Korea}                     
\centerline{$^{32}$CINVESTAV, Mexico City, Mexico}                            
\centerline{$^{33}$FOM-Institute NIKHEF and University of                     
                   Amsterdam/NIKHEF, Amsterdam, The Netherlands}              
\centerline{$^{34}$Radboud University Nijmegen/NIKHEF, Nijmegen, The          
                  Netherlands}                                                
\centerline{$^{35}$Joint Institute for Nuclear Research, Dubna, Russia}       
\centerline{$^{36}$Institute for Theoretical and Experimental Physics,        
                   Moscow, Russia}                                            
\centerline{$^{37}$Moscow State University, Moscow, Russia}                   
\centerline{$^{38}$Institute for High Energy Physics, Protvino, Russia}       
\centerline{$^{39}$Petersburg Nuclear Physics Institute,                      
                   St. Petersburg, Russia}                                    
\centerline{$^{40}$Lund University, Lund, Sweden, Royal Institute of          
                   Technology and Stockholm University, Stockholm,            
                   Sweden, and}                                               
\centerline{Uppsala University, Uppsala, Sweden}                              
\centerline{$^{41}$Physik Institut der Universit{\"a}t Z{\"u}rich,            
                   Z{\"u}rich, Switzerland}                                   
\centerline{$^{42}$Lancaster University, Lancaster, United Kingdom}           
\centerline{$^{43}$Imperial College, London, United Kingdom}                  
\centerline{$^{44}$University of Manchester, Manchester, United Kingdom}      
\centerline{$^{45}$University of Arizona, Tucson, Arizona 85721, USA}         
\centerline{$^{46}$Lawrence Berkeley National Laboratory and University of    
                   California, Berkeley, California 94720, USA}               
\centerline{$^{47}$California State University, Fresno, California 93740, USA}
\centerline{$^{48}$University of California, Riverside, California 92521, USA}
\centerline{$^{49}$Florida State University, Tallahassee, Florida 32306, USA} 
\centerline{$^{50}$Fermi National Accelerator Laboratory,                     
            Batavia, Illinois 60510, USA}                                     
\centerline{$^{51}$University of Illinois at Chicago,                         
            Chicago, Illinois 60607, USA}                                     
\centerline{$^{52}$Northern Illinois University, DeKalb, Illinois 60115, USA} 
\centerline{$^{53}$Northwestern University, Evanston, Illinois 60208, USA}    
\centerline{$^{54}$Indiana University, Bloomington, Indiana 47405, USA}       
\centerline{$^{55}$University of Notre Dame, Notre Dame, Indiana 46556, USA}  
\centerline{$^{56}$Purdue University Calumet, Hammond, Indiana 46323, USA}    
\centerline{$^{57}$Iowa State University, Ames, Iowa 50011, USA}              
\centerline{$^{58}$University of Kansas, Lawrence, Kansas 66045, USA}         
\centerline{$^{59}$Kansas State University, Manhattan, Kansas 66506, USA}     
\centerline{$^{60}$Louisiana Tech University, Ruston, Louisiana 71272, USA}   
\centerline{$^{61}$University of Maryland, College Park, Maryland 20742, USA} 
\centerline{$^{62}$Boston University, Boston, Massachusetts 02215, USA}       
\centerline{$^{63}$Northeastern University, Boston, Massachusetts 02115, USA} 
\centerline{$^{64}$University of Michigan, Ann Arbor, Michigan 48109, USA}    
\centerline{$^{65}$Michigan State University,                                 
            East Lansing, Michigan 48824, USA}                                
\centerline{$^{66}$University of Mississippi,                                 
            University, Mississippi 38677, USA}                               
\centerline{$^{67}$University of Nebraska, Lincoln, Nebraska 68588, USA}      
\centerline{$^{68}$Princeton University, Princeton, New Jersey 08544, USA}    
\centerline{$^{69}$State University of New York, Buffalo, New York 14260, USA}
\centerline{$^{70}$Columbia University, New York, New York 10027, USA}        
\centerline{$^{71}$University of Rochester, Rochester, New York 14627, USA}   
\centerline{$^{72}$State University of New York,                              
            Stony Brook, New York 11794, USA}                                 
\centerline{$^{73}$Brookhaven National Laboratory, Upton, New York 11973, USA}
\centerline{$^{74}$Langston University, Langston, Oklahoma 73050, USA}        
\centerline{$^{75}$University of Oklahoma, Norman, Oklahoma 73019, USA}       
\centerline{$^{76}$Oklahoma State University, Stillwater, Oklahoma 74078, USA}
\centerline{$^{77}$Brown University, Providence, Rhode Island 02912, USA}     
\centerline{$^{78}$University of Texas, Arlington, Texas 76019, USA}          
\centerline{$^{79}$Southern Methodist University, Dallas, Texas 75275, USA}   
\centerline{$^{80}$Rice University, Houston, Texas 77005, USA}                
\centerline{$^{81}$University of Virginia, Charlottesville,                   
            Virginia 22901, USA}                                              
\centerline{$^{82}$University of Washington, Seattle, Washington 98195, USA}  
}                                                                             

%% file: acknowledgement_paragraph_r2.tex
%
We thank the staffs at Fermilab and collaborating institutions, 
and acknowledge support from the 
DOE and NSF (USA);
CEA and CNRS/IN2P3 (France);
FASI, Rosatom and RFBR (Russia);
CAPES, CNPq, FAPERJ, FAPESP and FUNDUNESP (Brazil);
DAE and DST (India);
Colciencias (Colombia);
CONACyT (Mexico);
KRF and KOSEF (Korea);
CONICET and UBACyT (Argentina);
FOM (The Netherlands);
PPARC (United Kingdom);
MSMT (Czech Republic);
CRC Program, CFI, NSERC and WestGrid Project (Canada);
BMBF and DFG (Germany);
SFI (Ireland);
The Swedish Research Council (Sweden);
Research Corporation;
Alexander von Humboldt Foundation;
and the Marie Curie Program.